\documentclass[conference]{IEEEtran}
\usepackage{bbm}
\usepackage{mathrsfs}
\usepackage{amssymb,amsmath,float}
\usepackage[pdftex]{graphicx}
\ifCLASSINFOpdf
\else
\fi
\newcommand\ha{\hat a}
\newcommand\had{\hat a^\dag}
\newcommand\hb{\hat b}
\newcommand\hbd{\hat b^\dag}
\newcommand\hc{\hat c}
\newcommand\hcd{\hat c^\dag}
\newcommand\hd{\hat d}
\newcommand\hdd{\hat d^\dag}
\newcommand\hD{\hat D}
\newcommand\hE{\hat E}

\newcommand\hJ{\hat J}
\newcommand\hK{\hat K}
\newcommand\hL{\hat L}
\newcommand\var{{\rm var}}
\newcommand\cov{{\rm cov}}

\newcommand\cT{\mathcal{T}}

\newcommand\de{\delta}

\newcommand\ta{\tau}

\newcommand\om{\omega}

\newcommand\<{\langle}
\renewcommand\>{\rangle}

\newcommand\ie{\emph{i.e.}}

\newcommand\beq{\begin{equation}}
\newcommand\eeq{\end{equation}}
\newcommand\bea{\begin{eqnarray}}
\newcommand\eea{\end{eqnarray}}
\newcommand\bal{\begin{align}}
\newcommand\eal{\end{align}}

\newcommand\fr{\frac}

\newcommand\ap{\approx}

\renewcommand\bal{\mbox{\boldmath$\alpha$}}

\begin{document}
%
\title{Measuring the brightness of classical noise dominated light at the shot noise limit?}



%
\author{\IEEEauthorblockN{Richard Lieu\IEEEauthorrefmark{1}, and
T.W.B. Kibble\IEEEauthorrefmark{2},
\IEEEauthorblockA{\IEEEauthorrefmark{1}{\it Department of Physics, University of Alabama,
Huntsville, AL 35899.}\\}
\IEEEauthorblockA{\IEEEauthorrefmark{2}{\it Blackett Laboratory, Imperial College, London SW7~2AZ, U.K.}\\}}}



\maketitle

\begin{abstract}
A recent claim by Lieu et al that beam splitter intensity subtraction (or homodyne with one vacuum port) followed by high resolution sampling can lead to detection of brightness of thermal light at the shot noise limit is reexamined here.  We confirm the calculation of Zmuidzinas that the claim of Lieu et al was falsified by an incorrect assumption about the correlations in thermal noise.
\end{abstract}


%
\IEEEpeerreviewmaketitle

\section{Introduction}

Recently, 
\cite{lie14} proposed a method of improving the sensitivity of radio telescopes, based on using a 50:50 beam splitter and measuring the difference signal between the two output beam intensities.  The original motivation of \cite{lie14} was based upon the supposition that the intensity difference has shot noise fluctuations given by a simple Poisson process with its mean subtracted away, and a variance equal to the mean photon rate of the(presumed stationary) incident beam, although the statistics of the noise distribution presented there were derived from the quantum theory of chaotic light and the higher moments were found to be slightly different from Poisson.  Subsequently, it was pointed out by 
\cite{zmu14} that the higher moments of \cite{lie14} were erroneous, due to an invalid assumption about the absence of correlations between non-overlapping time intervals.  Below, we present an improved and simplified version of the calculation of \cite{lie14} that reaches the same conclusion as \cite{zmu14}, {\it viz.} although it is possible to use the split-beam technique to achieve essentially the same accuracy as that of a direct measurement of the incoming signal, it is much harder (if at all possible) to do significantly better.

\section{An Improved Calculation}

This is firstly an alternative calculation to the ones in \cite{lie14} and \cite{zmu14}) of the degree of accuracy attainable by a difference measurement on a split beam, and secondly a discussion of the effect in frequency space via a discrete Fourier transform.  The results are essentially the same as those of Zmuidzinas, \ie~the split-beam technique achieves more or less the same accuracy that one can get from a direct measurement of the brightness of the original beam.

Ahead of the formal treatment, it may be useful to seek a heuristic understanding of the difference between \cite{lie14} and \cite{zmu14}.  If light comprises only shot noise, the fluctuations in direct and homodyne measurements will in principle both look the same.  But, as is usually the case for chaotic light, there are classical phase noise fluctuations as well, and the shot noise will then exhibit a time dependent mean and variance as its amplitude varies together with the classical intensity noise in tandem.  According to \cite{zmu14} and the calculation here, this correlation between the shot noise variance and the classical noise intensity is not expected to be removed by the beam splitter, as illustrated in Figure 1.  \cite{lie14}, on the other hand, asserted that the appearance of the homodyne difference signal remains like simple shot noise, \ie~ either there was no such correlation in the incident beam to begin with, or the beam-splitter removed the effect.

\begin{figure}
\begin{center}
\includegraphics[angle=0,width=3.5in]{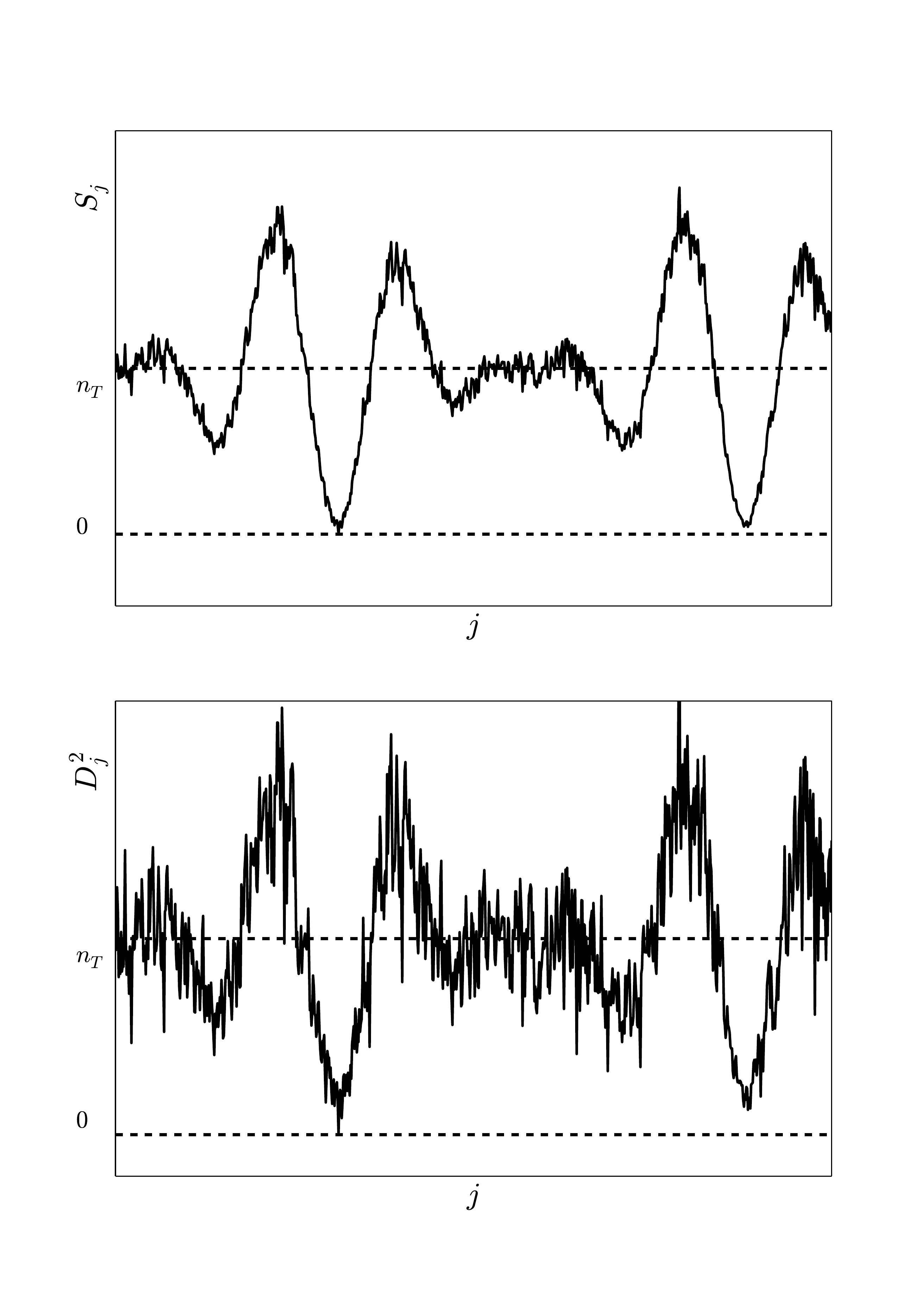}
\end{center}
\caption{Theoretically expected noise characteristics of stationary thermal radiation as viewed directly and via a beam-splitter.  Note the squared intensity difference $D_j^2$ still exhibits classical bunching noise that's correlated with the direct intensity time series $S_j$.  Additionally $D_j^2$ is also expected to have more shot noise.}

\end{figure}

\subsection{Direct measurements with incident beam}

So long as we are dealing with a narrow bandwidth, it is more convenient to work with the Fourier transforms of the annihilation and creation operators,
 \beq \ha(t) = \fr{1}{\sqrt{2\pi}} \int d\om\,\ha(\om)e^{-i\om t};
 \had(t) = \fr{1}{\sqrt{2\pi}} \int d\om\,\had(\om)e^{i\om t}. \eeq
They satisfy the commutation relations
 \beq [\ha(t), \ha(t')] = \de(t-t'). \label{comm} \eeq
For a chaotic beam with Gaussian frequency profile, centred on $\om_0$ and with bandwidth $1/\ta$, we have
 \beq \<\had(t)\ha(t')\> = n_0 f(t-t'), \label{corr} \eeq
where $f(t)$ is given by
 \beq f(t) = \fr{1}{\ta} e^{i\om_0 t}e^{-t^2/2\ta^2}. \eeq
Moreover, the intensity is simply $\om_0 \had(t)\ha(t)$, but in the narrow-band case, it is simpler to remove the factor of $\om_0$, and talk instead about
 \beq \hJ(t) = \had(t)\ha(t), \eeq
which represents the number of photons arriving per unit time.  It follows immediately that
 \beq \<\hJ(t)\> = \fr{n_0}{\ta}. \eeq

Next, we examine the covariance function
 \beq \cov(J(t), J(t')) = \<\hJ(t) \hJ(t')\> - \<\hJ\>^2. \label{covJ} \eeq
Now
 \bea \<\hJ(t) \hJ(t')\> \!\!\!&=&\!\!\! \<\had(t)\ha(t)\had(t')\ha(t')\>\notag\\
 \!\!\!&=&\!\!\! \<\had(t)\ha(t)\>\<\had(t')\ha(t')\> +\notag\\
  \!\!\!& &\!\!\! \<\had(t)\ha(t')\>\<\ha(t)\had(t')\>. \eea
The first term here clearly cancels the last term in (\ref{covJ}), so we find
 \beq \cov(J(t),J(t')) = n_0^2 |f(t-t')|^2 + \fr{n_0}{\ta}\de(t-t'), \label{corrJ}\eeq
where the final term comes from the commutator $[\ha(t),\had(t')]$.

If we define the average flux over a short time interval as
 \beq \hJ_T(t) = \fr{1}{T} \int_{t-T}^t dt'\,\hJ(t'), \eeq
then we find
 \beq \var(J_T(t)) =  \fr{1}{\ta T}\left[n_0^2 F\left(\fr{T}{\ta}\right) + n_0\right], \eeq
where
 \beq F\left(\fr{T}{\ta}\right) = \fr{\ta}{T} \int_{-T}^T dt\,(T-|t|)|f(t)|^2. \eeq
Note that for $T\ll \ta$, we may replace $f$ in the integrand by $1/\ta$, so $F(T/\ta)\ap T/\ta$.  The relative uncertainty in the measurement of $J_T$ is given by
 \beq \fr{\var(J_T(t))}{\<\hJ_T(t)\>^2} = \fr{\ta}{T} \left[F\left(\fr{T}{\ta}\right)
 + \fr{1}{n_0}\right], \eeq
or
 \beq \fr{\var(J_T(t))}{\<\hJ_T(t)\>^2} \ap 1+ \fr{\ta}{n_0 T},\quad\text{for}\quad T\ll\ta.
  \label{errorJ}\eeq

On the other hand, if we measure for a much longer time $\cT=NT$, we must use the limiting value of $f(x)$ for $x\gg 1$, namely $\sqrt{\pi}$.  So we have
 \beq \fr{\var(J_\cT(t))}{\<\hJ_\cT(t)\>^2} \ap \sqrt{\pi} \fr{\ta}{\cT}=\sqrt{\pi} \fr{\ta}{NT},
 \quad\text{for}\quad \cT\gg\ta.
 \label{errordirect}\eeq

\subsection{Difference signal for split beam}

In a 50:50 beam splitter, it is useful to consider a second input beam, which is in fact in its vacuum state.  Let us represent the annihilation and creation operators of that second input by $\hb(t), \hbd(t)$.  Then for the two output beams we have annihilation operators
 \beq \hc = \fr{1}{\sqrt{2}}(\ha+i\hb),\qquad \hd = \fr{1}{\sqrt{2}}(\ha-i\hb).
 \label{split} \eeq
Note that $\hc$ and $\hd$ each satisfy the commutation relations (\ref{comm}), together with (\ref{corr}) but with $n_0$ replaced by $n_0/2$.  Moreover, $\big[\hc,\hdd]=0$.

One might perhaps worry that using $\hb(t)$ rather than $\hb(\om)$, with the replacement of factors of $\om$ by $\om_0$, which is justified for the narrow-bandwidth case, might be inadmissible for the vacuum contribution.  However, if one retains the factors of $\om$, they will be converted to time derivatives that will ultimately act on other factors that are limited by bandwidth, and the leading contributions will be given quite accurately by the replacement of $\om$ by $\om_0$, so this is probably not a serious problem.

The quantity we are particularly interested in is the difference signal, the difference between the numbers of photons arriving in the two output channels.  This is given by
 \beq \hD(t) = \hcd(t)\hc(t) - \hdd(t)\hd(t). \eeq
Substituting from (\ref{split}) we see that this quantity may be written
 \beq \hD(t) = i\had(t)\hb(t)-i\hbd(t)\ha(t). \label{Dab} \eeq
Obviously, its expectation value is zero:
 \beq \<\hD(t)\>=0. \eeq
The factorization between $\ha$ and $\hb$ operators makes this a very convenient form to use.  For example, in computing the two-time function, we see that
 \beq \<\hD(t) \hD(t')\> = \<\had(t)\ha(t')\>\<\hb(t)\hbd(t')\>
 +\<\ha(t)\had(t')\>\<\hbd(t)\hb(t')\>, \eeq
and because the $b$ input is in its vacuum state, the second term vanishes, while in the first, $\<\hb(t)\hbd(t')\>=\de(t-t')$.  Thus we find
 \beq \cov(D(t),D(t')) = \<\hD(t) \hD(t')\> = \fr{n_0}{\ta}\de(t-t'). \eeq
So the measurement of the variance of $D$ provides a way of measuring $n_0$.

Of course, any measurement will take up a finite time interval.  We suppose that the total available time $\cT$ is divided up into $N$ small segments of duration $T$, and define the average flux in the $j$th interval as
 \beq \hD_j = \fr{1}{T}\int_{(j-1)T}^{jT} dt\,\hD(t), \eeq
where we assume $T\ll \ta$, so that
 \beq \var(D_j) = \fr{n_0}{T\ta}, \qquad \cov(D_j,D_k)=0,\ (j\ne k). \label{varD}\eeq

Now to estimate the accuracy of the measurement we can make, we need to compute the expectation value $\<\hD_j^2 \hD_k^2\>$.  However, for use later we consider the more general case
 \bea \<\hD_j \hD_k \hD_l \hD_m\> \!\!\!&=&\!\!\! \fr{1}{T^4} \int_{(j-1)T}^{jT} dt_1 \int_{(k-1)T}^{kT} dt_2\notag\\
\!\!\!& &\!\!\! \int_{(l-1)T}^{lT} dt_3 \int_{(m-1)T}^{mT} dt_4\notag\\
\!\!\!& &\!\!\! \<\hD(t_1)\hD(t_2)\hD(t_3)\hD(t_4)\>. \label{Dj4}\eea
When we substitute from (\ref{Dab}), each term in the expectation value can be written as a product of an expectation value of $\ha$ and $\had$ operators, and one of $\hb$ and $\hbd$ operators.  Moreover, the latter vanish if they have a $\hb$ on the right or a $\hbd$ on the left, and there must be equal numbers of each of the two terms in (\ref{Dab}) containing $\hb$ and $\hb^\dag$ operators.  So there are just two terms remaining:
 \bea \<\hD(t_1)\hD(t_2)\hD(t_3)\hD(t_4)\> \!\!\!&=&\!\!\!
 \<\had(t_1)\had(t_2)\ha(t_3)\ha(t_4)\>\notag\\
 \!\!\!& &\!\!\!\<\hb(t_1)\hb(t_2)\hbd(t_3)\hbd(t_4)\> +\notag\\
 \!\!\!& &\!\!\! \<\had(t_1)\ha(t_2)\had(t_3)\ha(t_4)\>\notag\\
 \!\!\!& &\!\!\! \<\hb(t_1)\hbd(t_2)\hb(t_3)\hbd(t_4)\>. \label{Dt4}\eea
Now, with the abbreviation $t_{jk}=t_j-t_k$,
 \beq \<\hb(t_1)\hb(t_2)\hbd(t_3)\hbd(t_4)\>=\de(t_{13})\de(t_{24})+\de(t_{14})\de(t_{23}), \eeq
while
 \beq \<\hb(t_1)\hbd(t_2)\hb(t_3)\hbd(t_4)\> = \de(t_{12})\de(t_{34}), \eeq
so clearly the result will only be nonzero when the indices $(j,k,l,m)$ are equal in pairs.

We also note that
 \beq \<\had(t_1)\had(t_2)\ha(t_3)\ha(t_4)\>=n_0^2[f(t_{13})f(t_{24})+f(t_{14})f(t_{23})]. \eeq
while
 \bea \<\had(t_1)\ha(t_2)\had(t_3)\ha(t_4)\> &=& n_0^2[f(t_{12})f(t_{34})+\notag\\
 \!\!\!& &\!\!\! f(t_{14})f(t_{32})] + \notag\\
\!\!\!& &\!\!\! n_0f(t_{14})\de(t_{23}). \eea

Putting these expressions together and substituting into (\ref{Dt4}), we find
 \bea &&\<\hD(t_1)\hD(t_2)\hD(t_3)\hD(t_4)\> = \notag\\
\!\!\!& &\!\!\! \de(t_{12})\de(t_{34})n_0^2\left(\fr{1}{\ta^2} + |f(t_{13})|^2\right) + \notag\\
\!\!\!& &\!\!\! \de(t_{13})\de(t_{24})n_0^2\left(\fr{1}{\ta^2}+ |f(t_{12})|^2\right) +\notag\\
\!\!\!& &\!\!\! \de(t_{14})\de(t_{23})n_0^2\left(\fr{1}{\ta^2}+ |f(t_{12})|^2\right) + \notag\\
\!\!\!& &\!\!\! \de(t_{12})\de(t_{23})\de(t_{34})\fr{n_0}{\ta}. \eea
Note the symmetry of this expression under permutations of $\{1,2,3,4\}$, which results from the fact that the different $\hD_j$ operators commute with each other.

Then, integrating over short time intervals, and assuming that $T\ll\ta$, we find
 \bea \<\hD_j \hD_k \hD_l \hD_m\> &=&
 \de_{jk}\de_{lm}\fr{n_0^2}{T^2\ta^2}(1+e^{-t^2_{jl}/\ta^2})+\notag\\
\!\!\!& &\!\!\! \de_{jl}\de_{km}\fr{n_0^2}{T^2\ta^2}(1+e^{-t^2_{jk}/\ta^2})+ \notag\\
\!\!\!& &\!\!\! \ \de_{jm}\de_{kl}\fr{n_0^2}{T^2\ta^2}(1+e^{-t^2_{jk}/\ta^2})+\notag\\
\!\!\!& &\!\!\! \de_{jk}\de_{kl}\de_{lm} \fr{n_0}{T^3\ta}, \label{Dj4} \eea
where $t_{jk}=(j-k)T$.

Now, to find the covariance of $D_j^2$ and $D_l^2$, we set $k=j$ and $m=l$, and remove the first of the seven terms in (\ref{Dj4}), which is cancelled by the product of expectation values.  This yields
 \beq \cov(D_j^2, D_l^2) = \fr{n_0^2}{T^2\ta^2}e^{-t_{jl}^2/\ta^2} + \de_{jl}\fr{4n_0^2}{T^2\ta^2}
 + \de_{jl}\fr{n_0}{T^3\ta}. \label{covD}\eeq
The first term alone gives the covariance when $j\ne l$.  For $j=l$ we find
 \beq \var(D_j^2) = 5\fr{n_0^2}{T^2\ta^2} + \fr{n_0}{T^3\ta}. \eeq
Thus the fractional error is given by
 \beq \fr{\var(D_j^2)}{\<\hD_j^2\>^2}= 5 + \fr{\ta}{n_0 T}. \eeq
This is comparable with (\ref{errorJ}) but larger (when $n_0T\gg \ta$) by a factor of 5.

Of course, as before we can do better by observing for a longer time.  In particular, we can form the sample mean
 \beq \overline{D^2} = \fr{1}{N}\sum_{j=1}^N D_j^2. \eeq
Clearly,
 \beq \var(\overline{D^2}) = \fr{1}{N^2}\sum_{j,l=1}^{N} \cov(D_j^2,D_l^2). \eeq
When we substitute from (\ref{covD}), in the first term, we can convert the sum over $j-l$ to a Gaussian integral:
 \beq \sum_j e^{-j^2T^2/\ta^2} \ap \fr{1}{T}\int dt\,e^{-t^2/\ta^2} = \fr{\sqrt{\pi}\ta}{T}. \eeq
Thus we obtain
 \beq \fr{\var(\overline{D^2})}{\<\hD_j^2\>^2} =
 \fr{1}{N}\left(\fr{\sqrt{\pi}\ta}{T}+4+\fr{\ta}{n_0 T}\right). \label{varDD}\eeq
The dominant term here, when $T\ll\ta\ll n_0T$ is the first.  This reproduces precisely the dominant term in the direct measurement error, (\ref{errordirect}).  So in this case we can do as well as the direct measurement, but unfortunately no better.

\subsection{Frequency domain measurements}

There is, however, another way of dealing with the information, in terms of the discrete finite Fourier transform of the signal.  Let us define
 \beq K_p = \fr{1}{\sqrt{N}}\sum_{j=1}^N D_j e^{-2\pi ipj/N},\qquad(p=0,\dots,N-1). \eeq
The frequency corresponding to $K_p$ is $\om_p = 2\pi p/NT$.  Note that $K_p$ is not real; in fact $K^*_p=K_{-p}\equiv K_{N-p}$.  Obviously, $\<\hK_p\> = 0$, and, from (\ref{varD}),
 \beq \<\hK_p\hK^\dag_p\> = \fr{n_0}{T\ta}, \qquad
 \<\hK_p\hK^\dag_q\>=0,\ (p\ne q), \label{KK}\eeq
very similar to the expressions for the variance and covariance of the $D_j$.  Thus for each $p$ the value of $|K_p|^2$ provides an estimate of $n_0$.

Next, we look at the uncertainty of these estimates.  To do so, we need the covariance of $|K_p|^2$ and $|K_q|^2$.  We start from (\ref{Dj4}) and apply a discrete Fourier transform to each of the variables.  This yields
 \bea \<\hK_p\hK_q \hK_r \hK_s\> \!\!\!&=&\!\!\! \de_{p+q+r+s} [\fr{n_0^2}{T^2\ta^2}
 \big(\de_{p+q}+G_{p+q}+\de_{p+r}+\notag\\
\!\!\!& &\!\!\! G_{p+r} + \de_{p+s}+G_{p+s}\big)+\fr{n_0}{NT^3\ta}], \eea
where $\de_p=1$ if $p\equiv 0 \mod N$, otherwise 0, and
 \beq G_p = \fr{1}{N}\sum_j e^{-2\pi ipj/N} e^{-j^2T^2/\ta^2} \ap
  \fr{\sqrt{\pi}\ta}{NT} e^{-(p\pi\ta/NT)^2}. \eeq
To find the covariance of $|K_p|^2$ and $|K_r|^2$ we have to set $q=-p$, $s=-r$, and again subtract the first term, to obtain
 \bea \cov(|K_p|^2, |K_r|^2) &=& \fr{n_0^2}{T^2\ta^2}(\de_{p+r}+\de_{p-r}
 +G_0+G_{p+r} \notag\\
 &+& G_{p-r}) + \fr{n_0}{NT^3\ta}. \eea
In particular, setting $p=r$ we have
 \bea \var(|K_p|^2) &=& \fr{n_0^2}{T^2\ta^2}\bigg(1+ \de_p + 2\fr{\sqrt{\pi}\ta}{NT}\notag\\
 &+& \fr{\sqrt{\pi}\ta}{NT} e^{-(2p\pi\ta/NT)^2}\bigg) + \fr{n_0}{NT^3\ta}. \label{covK} \eea
The fractional uncertainty measure is now
 \beq \fr{\var(|K_p|^2)}{\<\hK_p\hK^\dag_p\>^2} = 1+ \de_p + 2\fr{\sqrt{\pi}\ta}{NT}
 + \fr{\sqrt{\pi}\ta}{NT} e^{-(2p\pi\ta/NT)^2} + \fr{\ta}{n_0NT}. \eeq
This already shows that when $T\ll \ta\ll NT$, the variance of each $|K_p|^2$ is less than of each $D_j^2$.

As with the $D_j$, we can form the sample mean of these measurements,
 \beq \overline{|K|^2} = \fr{1}{N}\sum_{p=0}^{N-1} |K_p|^2, \eeq
so that
 \beq \var(\overline{|K|^2}) = \fr{1}{N^2}\sum_{p,r=0}^{N-1} \cov(|K_p|^2,|K_r|^2). \eeq
For the terms involving $G_p$ we can again convert the sum to an integral to give
 \beq \sum_p G_p = \fr{\sqrt{\pi}\ta}{NT}\sum e^{-(p\pi\ta/NT)^2} = 1. \eeq
Thus
 \beq \var(\overline{|K|^2}) = \fr{n_0^2}{NT^2\ta^2}\left(4+\fr{\sqrt{\pi}\ta}{T}\right)
 +\fr{n_0}{NT^3\ta}, \eeq
whence
 \beq \fr{\var(\overline{|K|^2})}{\<\hK_p\hK^\dag_p\>^2}=
 \fr{1}{N}\left(\fr{\sqrt{\pi}\ta}{T}+4+\fr{\ta}{n_0T}\right). \label{varKK} \eeq

\begin{figure}[!h]

  \centering
    \includegraphics[width=0.5\textwidth]{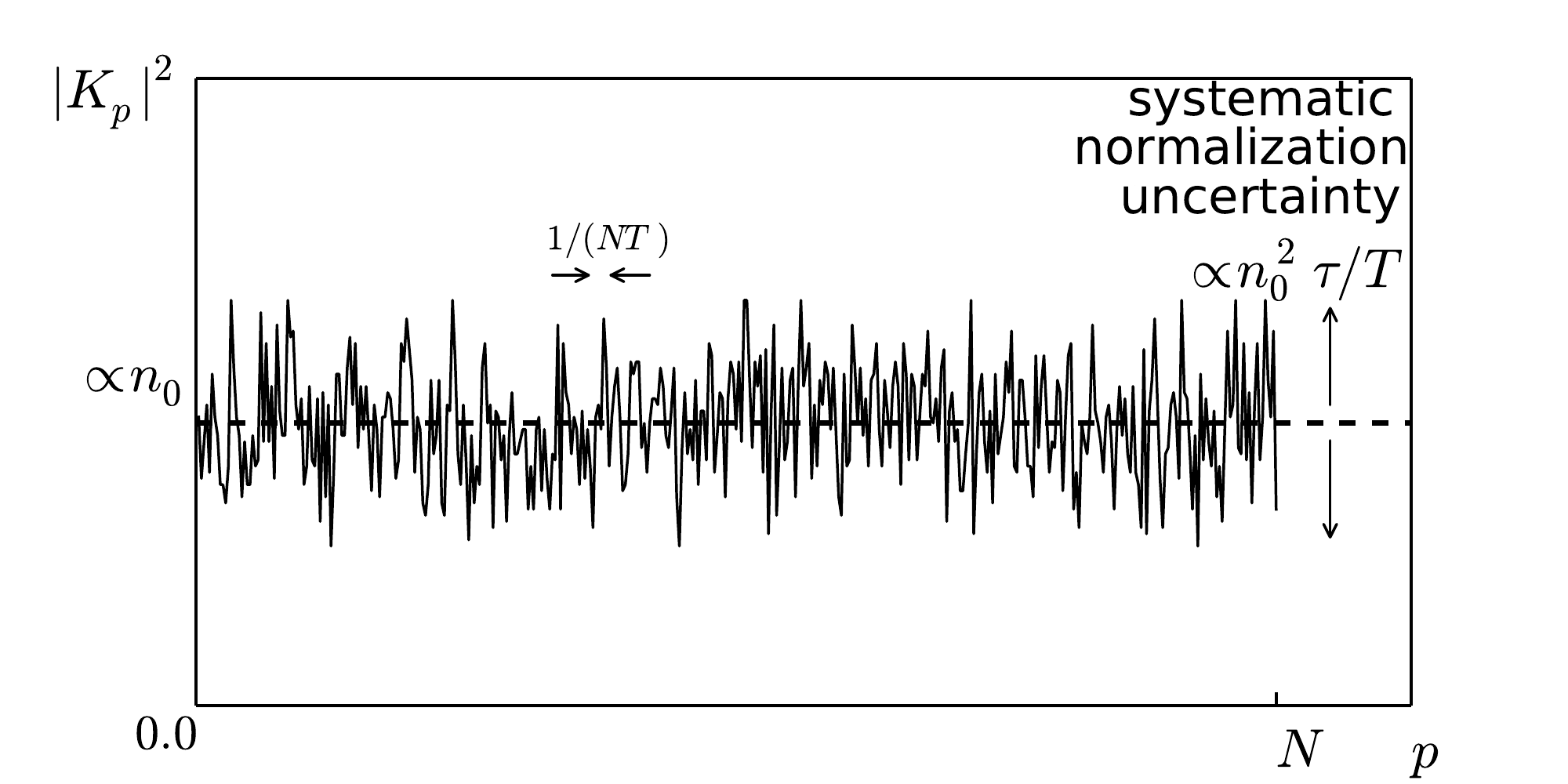}
    \label{homodyne}
 \caption{The Fourier power spectrum of the time series $D_j$ has both random and systematic uncertainties, with the latter in the form of a normalization error that shifts the entire (flat) spectrum up or down as measurements of the same thermal source are repeated.  The origin of this systematic effect, which for chaotic thermal light dominates the random noise when the spectral data are averaged over the full range of available frequencies, is photon bunching noise (also known as classical noise.}
\end{figure}

Note that this is \emph{precisely} the same as the result for $\overline{D^2}$.  This might seem surprising, given that the variance of each individual $|K_p^2|$ is less than that of each $D_j^2$.  However, it arises from the fact that there is some correlation between all the $|K_p^2|$, whereas for the $D_j^2$ it is limited to time differences less than about $\ta$.  The situation is depicted in figure (\ref{homodyne}).

\subsection{Fourier transforming $D^2$}

We could also envisage a different use of the finite discrete Fourier transform.  Let us define the quantity
$E_j = D_j^2$, which of course satisfies
 \beq \<\hE_j\>=\fr{n_0}{T\ta}. \eeq
Then we may define its discrete Fourier transform, say
 \beq L_p = \fr{1}{\sqrt{N}}\sum_{j=1}^N E_j e^{-2\pi ipj/N},\qquad(p=0,\dots,N-1). \eeq
It follows at once that
 \beq \<\hL_p\> = \sqrt{N}\fr{n_0}{T\ta}\de_p. \eeq
So in this case none of the Fourier components, except the DC (which apart from a normalization factor is the same thing as $\overline{D^2}$), can provide a measure of $n_0$.  So this does not seem a very profitable avenue of enquiry.

\section{Conclusion}

We revisited a recent claim \cite{lie14} that beam splitter intensity subtraction (or homodyne with one vacuum port) followed by high resolution sampling can lead to detection of brightness of thermal light at the shot noise limit.  We are able to confirm the result of \cite{zmu14} that the claim of \cite{lie14} was falsified by their incorrect assumption about thermal noise,

We should emphasize, however, that there has so far not been any experimental comparison of direct against homodyne brightness measurement of the {\it same} thermal state.  Thus the quantum field theoretic predictions remain to be verified in the laboratory.


\section{Acknowledgment}

We thank Jim Moran at Harvard CfA and Jonas Zmuidzinas at Caltech for helpful discussions.

\end{document}